\documentclass[11pt]{article}

\usepackage{osid1}
\usepackage{amsmath}
\usepackage{amssymb}
\usepackage{bm}

\title{Microscopic origin of Casimir-Polder forces}
\author{Stefan Yoshi
Buhmann\thanks{\hspace{1.5ex}\it e-mail:\ s.buhmann@tpi.uni-jena.de},
Hassan Safari, Dirk-Gunnar Welsch \\{\footnotesize\it
Theoretisch-Physikalisches Institut,
Friedrich-Schiller-Universit\"{a}t Jena,
Max-Wien-Platz 1, 07743 Jena, Germany}\\[2ex]
Ho Trung Dung \\{\footnotesize\it Institute of Physics, Academy of
Sciences and Technology, 1 Mac Dinh Chi Street, District 1,
Ho Chi Minh city, Vietnam}}

\begin{document}

\maketitle

\begin{abstract}
We establish a general relation between dispersion forces. First,
based on QED in causal media, leading-order perturbation theory is
used to express both the single-atom Casimir-Polder and the two-atom
van der Waals potentials in terms of the atomic polarizabilities and
the Green tensor for the body-assisted electromagnetic field. Endowed
with this geometry-independent framework, we then employ the Born
expansion of the Green tensor together with the Clausius-Mosotti
relation to prove that the macroscopic Casimir-Polder potential of an
atom in the presence of dielectric bodies is due to an infinite sum of
its microscopic many-atom van der Waals interactions with the atoms
comprising the bodies. This theorem holds for inhomogeneous,
dispersing, and absorbing bodies of arbitrary shapes and arbitrary
atomic composition on an arbitrary background of additional
magnetodielectric bodies.
\end{abstract}


\section{Introduction}
\label{Sec1}

The relation between the Casimir-Polder (CP) interaction of a single
atom with a body \cite{Lennard32,Casimir48} and its van der Waals
(vdW) interaction with the atoms comprising the body
\cite{Casimir48,London30} was first discussed in detail by Renne
\cite{Renne71}, with special emphasis on an atom interacting with a
dielectric half space. Approaching the problem from the microscopic
side and modelling all atoms by identical harmonic oscillators, he
showed that the sum of all many-atom vdW interactions between the
single atom and the body atoms corresponds to the result for the
CP potential derived earlier on the basis of a macroscopic approach
\cite{Lifshitz56}. Milonni and Lerner generalised this result to
bodies of arbitrary shapes \cite{Milonni92}. Using the Ewald-Oseen
extinction theorem which follows from the Clausius-Mosotti relation,
they demonstrated that the CP potential of an atom in the presence of
a nonabsorbing homogeneous dielectric body can be obtained by summing
over an infinite series of many-atom vdW potentials.

In this paper, we approach the problem from the macroscopic side and
under more general conditions. To that end, based on macroscopic QED
in linear, causal media (Sec.~\ref{Sec2}), we first consider the CP
interaction of an atom with a macroscopic body as well as the vdW
interaction of two atoms (Sec.~\ref{Sec3}), and then make use of the
Born expansion and the Clausius-Mosotti relation to establish the
microscopic origin of the CP potential (Sec.~\ref{Sec4}), followed by
a summary (Sec.~\ref{Sec5}).


\section{Atom-field interactions}
\label{Sec2}

According to the multipolar scheme, the Hamiltonian for a system of
neutral atoms and/or molecules --  briefly referred to as atoms in the
following -- interacting with the electromagnetic field in the
presence of dispersing and absorbing bodies is given by
\cite{Knoll01,Ho03,Buhmann04,Safari06}
\begin{equation}
\label{Eq14}
\hat{H}=\sum_A\hat{H}_A+\hat{H}_\mathrm{F}
 +\sum_A\hat{H}_{A\mathrm{F}},
\end{equation}
where the Hamiltonian
\begin{equation}
\label{Eq1}
\hat{H}_A=\sum_nE^n_A|n_A\rangle\langle n_A|
\end{equation}
governs the (unperturbed) internal dynamics of atom $A$,
\begin{equation}
\label{Eq3}
\hat{H}_{\mathrm{F}}
= \sum_{\lambda=e,m}\int\mathrm{d}^3r\,
 \int_0^\infty\mathrm{d}\omega\,\hbar\omega\,
 \hat{\mathbf{f}}_\lambda{\!^\dagger}(\mathbf{r},\omega)
 \!\cdot\!\hat{\mathbf{f}}_\lambda(\mathbf{r},\omega)
\end{equation}
is the Hamiltonian of the system composed of the electromagnetic field
and the magnetodielectric medium including dissipative interactions,
with $\hat{\mathbf{f}}_\lambda(\mathbf{r},\omega)$ and
$\hat{\mathbf{f}}_\lambda^\dagger(\mathbf{r},\omega)$ being the
dynamical variables of the system, which satisfy Bosonic commutation
relations, and the atom-field interaction in electric dipole
approximation is given by
\begin{equation}
\label{Eq6}
\hat{H}_{A\mathrm{F}}
= -\hat{\mathbf{d}}_A\!\cdot\!\hat{\mathbf{E}}(\mathbf{r}_A)
\end{equation}
($\hat{\mathbf{d}}_A$: atomic electric dipole moment, $\mathbf{r}_A$:
centre-of-mass position). In Eq.~(\ref{Eq6}),
\begin{equation}
\label{Eq8}
\hat{\mathbf{E}}(\mathbf{r})
=\sum_{\lambda=e,m}\int\mathrm{d}^3r'
 \int_0^\infty\mathrm{d}\omega\,
 \bm{G}_\lambda(\mathbf{r},\mathbf{r}',\omega)
 \!\cdot\!\hat{\mathbf{f}}_\lambda(\mathbf{r}',\omega)
 +\mathrm{H.c.}
\end{equation}
is the medium-assisted electric field expressed in terms of the
dynamical variables, with the quantities
\begin{align}
\label{Eq9}
\bm{G}_e(\mathbf{r},\mathbf{r}',\omega)
&= i\,\frac{\omega^2}{c^2}\,
 \bm{G}(\mathbf{r},\mathbf{r}',\omega)\,
 \sqrt{\frac{\hbar}{\pi\varepsilon_0}\,
 \mathrm{Im}\,\varepsilon(\mathbf{r}',\omega)}\,,\\
\label{Eq10}
\bm{G}_m(\mathbf{r},\mathbf{r}',\omega)
&= -i\,\frac{\omega}{c}\,
 \bm{G}(\mathbf{r},\mathbf{r}',\omega)\!\times\!
 \overleftarrow{\bm{\nabla}}_{\!\mathbf{r}'}
 \sqrt{\frac{\hbar}{\pi\varepsilon_0}\,
 \frac{\mathrm{Im}\,\mu(\mathbf{r}',\omega)}
 {|\mu(\mathbf{r}',\omega)|^2}}
\end{align}
being given in terms of the classical Green tensor
$\bm{G}(\mathbf{r},\mathbf{r}',\omega)$,
\begin{equation}
\label{Eq11}
\biggl[\bm{\nabla}\!\times\!\mu^{-1}(\mathbf{r},\omega)
 \bm{\nabla}\!\times
 -\frac{\omega^2}{c^2}\,\varepsilon(\mathbf{r},\omega)\biggr]
 \bm{G}(\mathbf{r},\mathbf{r}',\omega)
 =\bm{\delta}(\mathbf{r}-\mathbf{r}').
\end{equation}
The (macroscopic) permittivity $\varepsilon(\mathbf{r},\omega)$ and
permittivity $\mu(\mathbf{r},\omega)$ satisfy the Kra\-mers-Kronig
relations and the conditions
$\mathrm{Im}\,\varepsilon(\mathbf{r},\omega)$ $\!>$
$\!0$ and $\mathrm{Im}\,\mu(\mathbf{r},\omega)$ $\!>$ $\!0$ imposed
for absorbing media. Note that the Green tensor has the useful
properties \cite{Ho03}
\begin{gather}
\label{Eq12}
\bm{G}^\ast(\mathbf{r},\mathbf{r}',\omega)
 =\bm{G}(\mathbf{r},\mathbf{r}',-\omega^\ast),
 \qquad
\bm{G}(\mathbf{r},\mathbf{r}',\omega)
 =\bm{G}^{\mathrm T}(\mathbf{r}',\mathbf{r},\omega),\\
\label{Eq13}
\sum_{\lambda=e,m}\int\mathrm{d}^3 s\,
 \bm{G}_\lambda(\mathbf{r},\mathbf{s},\omega)\!\cdot\!
 \bm{G}^{\ast\mathrm{T}}_\lambda(\mathbf{r}',\mathbf{s},\omega)
 =\frac{\hbar\mu_0}{\pi}\,\omega^2
 \mathrm{Im}\,\bm{G}(\mathbf{r},\mathbf{r}',\omega).
\end{gather}


\section{Dispersion forces}
\label{Sec3}

Dispersion forces can be derived from the associated potentials $U$,
which are commonly identified as the position-dependent part of the
ground-state energy shift $\Delta E$ induced by the atom-field
coupling.


\subsection{The Casimir-Polder force}
\label{Sec3.1}

For a single atom $A$ the ground state of the system is given by
$|0\rangle$ $\!=$ $\!|0_A\rangle|\{0\}\rangle$ [$|0_A\rangle$, atomic
ground state; $\hat{f}_{\lambda i}(\mathbf{r},\omega)|\{0\}\rangle$
$\!=$ $\!0$], and the leading, second-order energy shift reads
\begin{equation}
\label{Eq15}
\Delta_2 E =\sideset{}{'}\sum_{I}
\frac{\langle 0|\hat{H}_{A\mathrm{F}}|I\rangle
 \langle I|\hat{H}_{A\mathrm{F}}|0\rangle}{E_0-E_I}\,.
\end{equation}
Note that the primed sum includes (principal-value)
integrals over the continuous degrees of freedom. Inspection of
Eq.~(\ref{Eq6}) reveals that only intermediate states of the type
$|I\rangle$ $\!=$ $\!|k_A\rangle 
\hat{f}^\dagger_{\lambda i}(\mathbf{r},\omega)|\{0\}\rangle$
contribute. Substituting the respective matrix elements
\begin{equation}
\label{Eq16}
\langle 0_A|\langle\{0\}|
 -\hat{\mathbf{d}}_A\!\cdot\!\hat{\mathbf{E}}(\mathbf{r}_A)
 \hat{f}^\dagger_{\lambda i}(\mathbf{r},\omega)|\{0\}\rangle
 |k_A\rangle
 =-\bigl[\mathbf{d}_A^{0k}\!\cdot\!
 \bm{G}_\lambda(\mathbf{r}_A,\mathbf{r},\omega)\bigr]_i
\end{equation}
[$\mathbf{d}_A^{0k}$ $\!=$ 
$\!\langle 0_A|\hat{\mathbf{d}}_A|k_A\rangle$] and energy denominators
$E_0-E_I$ $\!=$ $-\hbar(\omega_A^k+\omega)$ [$\omega_A^k$ $\!=$
$\!(E_A^k-E_A^0)/\hbar$] into Eq.~(\ref{Eq15}), using
Eq.~(\ref{Eq13}), and separating the Green tensor into its bulk
and scattering parts according to
\begin{equation}
\label{Eq18}
\bm{G}(\mathbf{r}_A,\mathbf{r}_A,\omega)
 =\bm{G}^{(0)}(\mathbf{r}_A,\mathbf{r}_A,\omega)
 +\bm{G}^{(1)}(\mathbf{r}_A,\mathbf{r}_A,\omega),
\end{equation}
one arrives, after some calculation, at [$\Delta_2E$ $\!\mapsto$
$\!U_A(\mathbf{r}_A)$] \cite{Buhmann04}
\begin{equation}
\label{Eq19}
U_A(\mathbf{r}_A)=\frac{\hbar\mu_0}{2\pi}
 \int_0^\infty\mathrm{d}u\,u^2\alpha_A(iu)
 \,\mathrm{Tr}\,
 \bm{G}^{(1)}(\mathbf{r}_A,\mathbf{r}_A,iu),
\end{equation}
where, for simplicity, an isotropic atomic ground-state polarizability
is considered,
\begin{equation}
\label{Eq20}
\alpha_A(\omega)=\lim_{\epsilon\to 0}\frac{2}{3\hbar}
 \sum_k\frac{\omega_A^k|\mathbf{d}_A^{0n}|^2}
 {(\omega_A^k)^2-\omega^2-i\omega\epsilon}\,.
\end{equation}
The potential (\ref{Eq19}) implies the CP force
($\bm{\nabla}_{\!\!A}$ $\!\equiv$ $\!\bm{\nabla}_{\!\mathbf{r}_A}$)
\begin{equation}
\label{Eq21}
\mathbf{F}_A=-\bm{\nabla}_{\!\!A}U_A(\mathbf{r}_A)
\end{equation}
on a ground-state atom $A$ due to the presence of an arbitrary
arrangement of dispersing and absorbing bodies [which are accounted
for by $\bm{G}^{(1)}(\mathbf{r}_A,\mathbf{r}_A,iu)$].

To illustrate the use of Eq.~(\ref{Eq19}), consider an atom placed at
distance $r_A$ from the centre of a small, homogeneous,
magnetodielectric sphere of radius $R$ $\!\ll$ $\!r_A$. Substituting
the respective Green tensor \cite{Li94} into Eq.~(\ref{Eq19}) and
retaining only the leading-order term in $R/r_A$ (cf.\
Ref.~\cite{Buhmann04b}), one obtains
\begin{align}
\label{Eq22}
U_A(r_A) = &-\frac{\hbar}{32\pi^3\varepsilon_0^2r_A^6}
 \int_0^\infty\mathrm{d}u\,\alpha_A(iu)\bigl[
 g_{ee}(ur_\mathrm{A}/c)\alpha_\odot^e(iu)
 \nonumber\\
& -(ur_A)^2/c^4g_{em}(ur_A/c)\alpha_\odot^m(iu)
 \bigr],\\
\label{Eq23}
g_{ee}(x) = &\; 2e^{-2x}(3+6x+5x^2+2x^3+x^4),\\
\label{Eq24}
g_{em}(x) = &\; 2e^{-2x}(1+2x+x^2),
\end{align}
with the electric and magnetic polarizabilities of the sphere
being given by \cite{Jackson98}
\begin{equation}
\label{Eq25}
\alpha_\odot^e(\omega)=4\pi\varepsilon_0R^3
 \,\frac{\varepsilon(\omega)-1}{\varepsilon(\omega)+2}\,,\qquad
\alpha_\odot^m(\omega)=\frac{4\pi R^3}{\mu_0}\,
 \frac{\mu(\omega)-1}{\mu(\omega)+2}\,.
\end{equation}
In the nonretarded limit, where the atom-sphere separation is
much larger than the characteristic transition wavelengths of the atom
and the sphere medium, one may approximate
$g_{ee}(ur_\mathrm{A}/c)$ $\!\simeq$ $g_ {ee}(0)$,
$g_{em}(ur_\mathrm{A}/c)$ $\!\simeq$ $g_{em}(0)$, so Eq.~(\ref{Eq22})
reduces to
\begin{equation}
\label{Eq27}
U_A(r_A) = -\frac{3\hbar}{16\pi^3\varepsilon_0^2r_A^6}
 \int_0^\infty\mathrm{d}u\,\alpha_A(iu)
 \alpha_\odot^e(iu)
\end{equation}
and
\begin{equation}
\label{Eq28}
U_A(r_A)
=  \frac{\hbar\mu_0}{16\pi^3\varepsilon_0r_A^4}
 \int_0^\infty\mathrm{d}u\,\Bigl(\frac{u}{c}\Bigr)^2
 \alpha_A(iu)\alpha_\odot^m(iu)
\end{equation}
for electric and magnetic spheres, respectively, while in the opposite
retarded limit the approximations $\alpha_A(iu)$ $\!\simeq$
$\!\alpha_A(0)$, $\alpha_\odot^e(iu)$ $\!\simeq$
$\!\alpha_\odot^e(0)$, $\alpha_\odot^m(iu)$ $\!\simeq$
$\!\alpha_\odot^m(0)$ lead to
\begin{eqnarray}
\label{Eq29}
U_A(r_A)=-\frac{\hbar c\alpha_A(0)
 \bigl[23\alpha_\odot^e(0)
 -7\alpha_\odot^m(0)/c^2\bigr]}
 {64\pi^3\varepsilon_0^2r_\mathrm{A}^7}\,.
\end{eqnarray}


\subsection{The two-atom van der Waals force}
\label{Sec3.2}

To calculate the vdW interaction of two atoms $A$ and $B$ in the
presence of dispersing and absorbing bodies, we start from the ground
state $|0\rangle$ $\!=$ $\!|0_A0_B\rangle|\{0\}\rangle$ and consider
those contributions to the energy shift that depend on the positions
of both atoms. The leading contributions are hence contained in the
fourth-order perturbative shift
 \begin{equation}
\label{Eq30}
\Delta_4E=\sideset{}{'}\sum_{I,II,III}
\frac{\langle 0|\hat{H}_\mathrm{int}|I\rangle
 \langle I|\hat{H}_\mathrm{int}|II\rangle
 \langle II|\hat{H}_\mathrm{int}|III\rangle
 \langle III|\hat{H}_\mathrm{int}|0\rangle}
 {(E_0-E_I)(E_0-E_{II})(E_0-E_{III})}\,,
\end{equation}
where $\hat{H}_\mathrm{int}$ $\!=$ $\!\hat{H}_{A\mathrm{F}}$ $\!+$
$\!\hat{H}_{B\mathrm{F}}$. A typical set of possible intermediate
states is given by
\begin{gather}
|I_{(1)}\rangle=|k_A0_B\rangle
 \hat{f}_{\lambda i}^\dagger(\mathbf{r},\omega)|\{0\}\rangle,\qquad
|III_{(1)}\rangle=|0_Al_B\rangle
 \hat{f}_{\lambda'''i'''}^\dagger(\mathbf{r}''',\omega''')
 |\{0\}\rangle,\nonumber\\
\label{Eq31}
|II_{(1)}\rangle=|0_A0_B\rangle{\textstyle\frac{1}{\sqrt{2}}}
 \hat{f}_{\lambda' i'}^\dagger(\mathbf{r}',\omega')
 \hat{f}_{\lambda'' i''}^\dagger(\mathbf{r}'',\omega'')
 |\{0\}\rangle.
\end{gather}
Upon using Eq.~(\ref{Eq16}) as well as
\begin{align}
\label{Eq32}
&-\langle k_A0_B|\langle\{0\}|
 \hat{f}_{\lambda i}(\mathbf{r},\omega)
 \hat{\mathbf{d}}_A\!\cdot\!\hat{\mathbf{E}}(\mathbf{r}_A)
 |0_A0_B\rangle
 \hat{f}_{\lambda' i'}^\dagger(\mathbf{r}',\omega')
 \hat{f}_{\lambda'' i''}^\dagger(\mathbf{r}'',\omega'')
 |\{0\}\rangle\nonumber\\
&\hspace{4ex}=-\Bigl\{\bigl[\mathbf{d}_A^{k0}\!\cdot\!
 \bm{G}_{\lambda'}(\mathbf{r}_A,\mathbf{r}',\omega')\bigr]_{i'}
 \delta_{\lambda\lambda^{\prime\prime}}\delta_{ii^{\prime\prime}}
 \delta(\mathbf{r} -\mathbf{r}^{\prime\prime})
 \delta(\omega-\omega^{\prime\prime})\nonumber\\
&\hspace{4ex}\quad\,+\bigl[\mathbf{d}_A^{k0}\!\cdot\!
 \bm{G}_{\lambda^{\prime\prime}}(\mathbf{r}_A,
 \mathbf{r}^{\prime\prime},\omega^{\prime\prime})
 \bigr]_{i^{\prime\prime}}
 \delta_{\lambda\lambda'}\delta_{ii'}
 \delta(\mathbf{r} -\mathbf{r}')
 \delta(\omega-\omega')\Bigr\}
\end{align}
and recalling Eq.~(\ref{Eq13}), substitution of the intermediate
states (\ref{Eq31}) into Eq.~(\ref{Eq30}) leads, after some
calculation, to
\begin{align}
\label{Eq33}
\Delta_4^{(1)}E
=&-\frac{\mu_0^2}{\hbar\pi^2}\sum_{k,l}
 \int_0^\infty\mathrm{d}\omega\omega^2\,
 \mathbf{d}_A^{0k}\!\cdot\!
 \mathrm{Im}\,\bm{G}(\mathbf{r}_A,\mathbf{r}_B,\omega)
 \!\cdot\!\mathbf{d}_B^{0l}
 \int_0^\infty\mathrm{d}\omega'\omega'^{2}
 \nonumber\\
&\times\biggl[\frac{\mathbf{d}_A^{0k}\!\cdot\!
 \mathrm{Im}\,\bm{G}(\mathbf{r}_A,\mathbf{r}_B,\omega')
 \!\cdot\!\mathbf{d}_B^{0l}}
 {(\omega_A^k\!+\!\omega)(\omega\!+\!\omega')
 (\omega_B^l\!+\!\omega)}
 +\frac{\mathbf{d}_A^{0k}\!\cdot\!
 \mathrm{Im}\,\bm{G}(\mathbf{r}_A,\mathbf{r}_B,\omega')
 \!\cdot\!\mathbf{d}_B^{0l}}
 {(\omega_A^k\!+\!\omega)(\omega\!+\!\omega')
 (\omega_B^l\!+\!\omega')}\biggr],
\end{align}
where we have assumed real dipole-matrix elements. Under this
assumption, the various two-atom contributions $\Delta_4^{(j)}E$
to the energy shift $\Delta_4E$ only differ by the denominators in the
square brackets of Eq.~(\ref{Eq33}), and hence after a lengthy
calculation one may show that [$\Delta_4E$ $\!\mapsto$
$\!U_{AB}(\mathbf{r}_A,\mathbf{r}_B)$] \cite{Safari06,Craig84}
\begin{equation}
\label{Eq37}
U_{AB}(\mathbf{r}_A,\mathbf{r}_\mathrm{B})
 =-\frac{\hbar\mu_0^2}{2\pi}\!\int_0^\infty\!\!\mathrm{d}u\,u^4
 \alpha_A(iu)\alpha_B(iu)
 \mathrm{Tr}\bigl[
 \bm{G}(\mathbf{r}_A,\mathbf{r}_B,iu)\!\cdot\!
 \bm{G}(\mathbf{r}_B,\mathbf{r}_A,iu)\bigr].
\end{equation}
From the two-atom potential (\ref{Eq37}) one can calculate the vdW
force on atom $A$($B$) due to atom $B$($A$) in the presence of
arbitrary dispersing and absorbing magnetodielectric bodies according to
 \begin{equation}
\label{Eq38}
\mathbf{F}_{A(B)}
 =-\bm{\nabla}_{\!\!A(B)}
 U_{AB}(\mathbf{r}_\mathrm{A},\mathbf{r}_\mathrm{B}).
\end{equation}

As a simple example, consider two atoms embedded in bulk
magnetodielectric material. Substitution of the respective Green
tensor \cite{Knoll01} into Eq.~(\ref{Eq37}) leads to
\begin{eqnarray}
\label{Eq39}
U_{AB}(r_{AB})=-\frac{\hbar}{32\pi^3\varepsilon_0^2r_{AB}^6}
 \int_0^\infty\mathrm{d}u\,\alpha_A(iu)\alpha_B(iu)\,
 \frac{g_{ee}[n(iu)ur_\mathrm{A}/c]}{\varepsilon^2(iu)}\,,
\end{eqnarray}
$r_{AB}$ $\!\equiv$ $\!|\mathbf{r}_A-\mathbf{r}_B|$, $n(\omega)$
$\!=$ $\!\sqrt{\varepsilon(\omega)\mu(\omega)}$, recall
Eq.~(\ref{Eq23}), which reduces to
\begin{align}
\label{Eq40}
&U(r_{AB}) = -\frac{3\hbar}{16\pi^3\varepsilon_0^2r_{AB}^6}
 \int_0^\infty\mathrm{d}u\,
 \frac{\alpha_A(iu)\alpha_B(iu)}{\varepsilon^2(iu)}\,,\\
\label{Eq41}
&U(r_{AB}) = -\frac{23\hbar c\alpha_A(0)\alpha_B(0)}
 {64\pi^3\varepsilon_0^2\varepsilon^2(0)n(0)r_{AB}^7}
\end{align}
for nonretarded and retarded interatomic separations, respectively.
Equations (\ref{Eq39})--(\ref{Eq41}) show that the presence of a
medium leads to a reduction of the potential w.r.t.\ its well-known
free-space value \cite{Casimir48}, while comparison of
Eqs.~(\ref{Eq22}) and (\ref{Eq39}) reveals that in free space the
dispersion interaction of an atom with a small dielectric sphere has
the same form as that of two atoms.


\section{Relation between dispersion forces}
\label{Sec4}

We now turn to the question how the CP interaction of a single atom
with dielectric bodies can be related to its many-atom vdW
interactions with the atoms comprising the bodies. For simplicity, we
will speak of a single dielectric body in the following. We assume the
dielectric body to be given by $\chi(\mathbf{r},\omega)$, and we allow
for the presence of an arbitrary magnetodielectric background of
additional bodies characterised by
$\overline{\varepsilon}(\mathbf{r},\omega)$ and
$\mu(\mathbf{r},\omega)$, such that
\begin{equation}
\label{Eq42}
\varepsilon(\mathbf{r},\omega)
 =\overline{\varepsilon}(\mathbf{r},\omega)+\chi(\mathbf{r},\omega).
\end{equation}
The Green tensor corresponding to this scenario can formally be
written as a Born series \cite{Buhmann05}
\begin{align}
\label{Eq43}
\bm{G}(\mathbf{r},\mathbf{r}',\omega)
 =&\;\overline{\bm{G}}(\mathbf{r},\mathbf{r}',\omega)
 +\sum_{k=1}^\infty\Bigl(\frac{\omega}{c}\Bigr)^{2k}
 \Biggl[\prod_{i=1}^k\int\mathrm{d}^3s_i\,
 \chi(\mathbf{s}_i,\omega)\Biggr]\nonumber\\
&\hspace{4ex}\times\,
 \overline{\bm{G}}(\mathbf{r},\mathbf{s}_1,\omega)\!\cdot\!
 \overline{\bm{G}}(\mathbf{s}_1,\mathbf{s}_2,\omega)\!\cdots\!
 \overline{\bm{G}}(\mathbf{s}_k,\mathbf{r}',\omega),
\end{align}
where $\overline{\bm{G}}$ is the Green tensor corresponding to the
magnetodielectric background. Substituting Eq.~(\ref{Eq43}) into
Eq.~(\ref{Eq19}), the CP potential can be written as
\begin{equation}
\label{Eq45}
U_A(\mathbf{r}_A)=\overline{U}_A(\mathbf{r}_A)
 +\sum_{k=1}^\infty\Delta_kU_A(\mathbf{r}_A),
\end{equation}
where
\begin{equation}
\label{Eq46}
\overline{U}_A(\mathbf{r}_A)
 =\frac{\hbar\mu_0}{2\pi}\int_0^\infty\mathrm{d}u\,u^2\alpha_A(iu)
 \,\mathrm{Tr}\,\overline{\bm{G}}^{(1)}(\mathbf{r}_A,\mathbf{r}_A,iu)
\end{equation}
is the CP potential due to the magnetodielectric background, which is
not of further interest here, and
\begin{eqnarray}
\label{Eq47}
\Delta_kU_A(\mathbf{r}_A)
&=&\frac{(-1)^k\hbar\mu_0}{2\pi c^{2k}}
 \int_0^{\infty}\mathrm{d}u\,u^{2k+2}\alpha_A(iu)
 \Biggl[\prod_{i=1}^k\int\mathrm{d}^3s_i\,
 \chi(\mathbf{s}_i,iu)\Biggr]
 \nonumber\\
&&\times\mathrm{Tr}\bigl[
 \overline{\bm{G}}(\mathbf{r}_A,\mathbf{s}_1,iu)\!\cdot\!
\overline{\bm{G}}(\mathbf{s}_1,\mathbf{s}_2,iu)\!\cdots\!
\overline{\bm{G}}(\mathbf{s}_k,\mathbf{r}_A,iu)\bigr]
\end{eqnarray}
is the contribution to the potential that is of $k$th order in
$\chi$.

Assuming the dielectric body described by $\chi$ to be comprised of
atoms of polarizabilities $\alpha_B(\omega)$ and number densities
$n_B(\mathbf{r})$, the gap between the macroscopic and microscopic
descriptions can be bridged by means of the Clausius-Mosotti law
\begin{equation}
\label{Eq48}
\chi(\mathbf{r},\omega)
 =\frac{\sum_B n_B(\mathbf{r})\alpha_B(\omega)/\varepsilon_0}
 {1-\sum_Cn_C(\mathbf{r})\alpha_C(\omega)/(3\varepsilon_0)}\,.
\end{equation}
Note that since $\chi$ is the Fourier transform of a linear response
function, it must satisfy the condition $\chi(\mathbf{r},0)$ $\!>$
$\chi(\mathbf{r},iu)$ $\!>$ $\!0$ for $u$ $\!>$ $\!0$, which implies
that the inequality
\begin{equation}
\label{Eq49}
\frac{1}{3\varepsilon_0}\sum_Bn_B(\mathbf{r})\alpha_B(0) < 1
\end{equation}
must hold.
Substituting Eq.~(\ref{Eq48}) into Eq.~(\ref{Eq47}) and splitting off
the singular part of the Green tensor according to
\begin{equation}
\label{Eq50}
\overline{\bm{G}}(\mathbf{r},\mathbf{r}',iu)
 =\frac{1}{3}\Bigl(\frac{c}{u}\Bigr)^2
 \bm{\delta}(\mathbf{r}-\mathbf{r}')
 +\overline{\bm{H}}(\mathbf{r},\mathbf{r}',iu),
\end{equation}
one obtains
\begin{equation}
\label{Eq51}
\Delta_kU_A(\mathbf{r}_A)=\sum_{l=1}^k\Delta_k^lU_A(\mathbf{r}_A),
\end{equation}
where
\begin{multline}
\label{Eq52}
\Delta_k^lU_A(\mathbf{r}_A)
 =\sum_{\substack{\eta_1\ge 0,\ldots,\eta_l\ge 0\\
 \eta_1+\ldots+\eta_l=k-l}}
 \int_0^\infty\mathrm{d}u\Biggl[\prod_{i=1}^l\int\mathrm{d}^3s_i\,
 \frac{\sum_{B_i}\,n_{B_i}(\mathbf{s}_i)
 q^{\eta_i}(\mathbf{s}_i,iu)}
 {1-\sum_{C_i}n_{C_i}(\mathbf{s}_i)
 \alpha_{C_i}(iu)/(3\varepsilon_0)}
 \Biggr]\\
\times\, V_{AB_1\ldots B_l}
 (\mathbf{r}_A,\mathbf{s}_1,\ldots,\mathbf{s}_l,iu)
\end{multline}
with
\begin{multline}
\label{Eq53}
V_{A_1\ldots A_j}(\mathbf{r}_1,\ldots,\mathbf{r}_j)
 \equiv\int_0^\infty\mathrm{d}u\,
 V_{A_1\ldots A_j}(\mathbf{r}_1,\ldots,\mathbf{r}_j,iu)
 =\frac{(-1)^{j-1}\hbar\mu_0^j}
 {2\pi}\int_0^\infty\mathrm{d}u\,u^{2j}\\
\times\,\alpha_{A_1}(iu)\cdots\alpha_{A_j}(iu)
 \mathrm{Tr}\bigl[
 \overline{\bm{H}}(\mathbf{r}_1,\mathbf{r}_2,iu)\cdots
 \overline{\bm{H}}(\mathbf{r}_j,\mathbf{r}_1,iu)\bigr]
\end{multline}
denotes the sum of all $l$-atom terms that are of order $k$ in
$\chi$, and each power of the factor
\begin{equation}
\label{Eq54}
q(\mathbf{r},iu)=
 -\frac{\sum_Bn_B(\mathbf{r})\alpha_B(iu)/(3\varepsilon_0)}
 {1-\sum_Cn_C(\mathbf{r})\alpha_C(iu)/(3\varepsilon_0)}
\end{equation}
is due to the integration of one term containing
$\delta(\mathbf{s}_i-\mathbf{s}_{i+1})$. Summing Eq.~(\ref{Eq51})
over $k$, one may rearrange the double sum as follows:
\begin{gather}
\label{Eq55}
\sum_{k=1}^\infty\Delta_kU_A(\mathbf{r}_A)
=\sum_{k=1}^\infty\sum_{l=1}^k\Delta_k^lU_A(\mathbf{r}_A)
=\sum_{l=1}^\infty\Delta^lU_A(\mathbf{r}_A),\\
\label{Eq56}
\Delta^lU(\mathbf{r}_A)
 =\Biggl[\prod_{i=1}^l\int\mathrm{d}^3s_i\,
 \sum_{B_i}\,n_{B_i}(\mathbf{s}_i)\Biggr]V_{AB_1\ldots B_l}
 (\mathbf{r}_A,\mathbf{s}_1,\ldots,\mathbf{s}_l),
\end{gather}
where we have performed the geometric sums $\sum_{\eta=0}^\infty
q^\eta$ by means of Eq.~(\ref{Eq54}). Note that the convergence of
these sums requires $|q(\mathbf{r},iu)|$ $\!<$ $\!1$, which by means
of Eq.~(\ref{Eq49}) is equivalent to
\begin{equation}
\label{Eq58}
\frac{2}{3\varepsilon_0}\sum_Bn_B(\mathbf{r})\alpha_B(0)<1.
\end{equation}

Finally, we symmetrize the many-atom terms by introducing the
symmetrization operator
\begin{equation}
\label{Eq59}
\mathcal{S}f(\mathbf{r}_1,\ldots,\mathbf{r}_j)
 \equiv\frac{1}{(2-\delta_{2j})j}\sum_{\pi\in P(j)}
 f(\mathbf{r}_{\pi(1)},\ldots,\mathbf{r}_{\pi(j)}),
\end{equation}
where $P(j)$ denotes the permutation group of the numbers
$1,\ldots,j$. From the cyclic property of the trace together
with Eq.~(\ref{Eq12}) it follows that
\begin{equation}
\label{Eq60}
\mathrm{Tr}\bigl[
 \overline{\bm{H}}(\mathbf{r}_1,\mathbf{r}_2,\omega)\cdots
 \overline{\bm{H}}(\mathbf{r}_j,\mathbf{r}_1,\omega)\bigr]
 =\mathrm{Tr}\bigl[
 \overline{\bm{H}}
 (\mathbf{r}_{\pi(1)},\mathbf{r}_{\pi(2)},\omega)\cdots
 \overline{\bm{H}}
 (\mathbf{r}_{\pi(j)},\mathbf{r}_{\pi(1)},\omega)\bigr],
\end{equation}
if $\pi$ is either a cyclic permutation or the reverse of a cyclic
permutation. Thus the sum on the r.h.s.\ of Eq.~(\ref{Eq59}) contains
classes of $(2-\delta_{2j})j$ terms that give the same result. By
forming a set $\overline{P}(j)$ $\!\varsubsetneq$ $\!P(j)$ containing
exactly one representative of each class, we can remove this
redundancy, leading to
\begin{align}
\label{Eq61}
&\mathcal{S}\mathrm{Tr}\bigl[
 \overline{\bm{H}}(\mathbf{r}_1,\mathbf{r}_2,\omega)\cdots
 \overline{\bm{H}}(\mathbf{r}_j,\mathbf{r}_1,\omega)\bigr]
 \nonumber\\
&\qquad=\sum_{\pi\in \overline{P}(j)}\!\!
 \mathrm{Tr}\bigl[
 \overline{\bm{H}}
 (\mathbf{r}_{\pi(1)},\mathbf{r}_{\pi(2)},\omega)\cdots
 \overline{\bm{H}}
 (\mathbf{r}_{\pi(j)},\mathbf{r}_{\pi(1)},\omega)\bigr].
\end{align}
Introducing the factor $1/l!$ in Eq.~(\ref{Eq56}) and summing over
all $l!$ possible ways of renaming the variables $\mathbf{s}_i$ and
$B_i$, the representative of each class in Eq.~(\ref{Eq61}) is
generated exactly twice (only once for $j$ $\!=$ $\!l\!+\!1$
$\!=$ $\!2$), so that the CP potential of atom $A$ due to the
dielectric body $\chi$ can be written as
\begin{equation}
\label{Eq62}
U_A(\mathbf{r}_A)=\sum_{l=1}^\infty\frac{1}{l!}\Biggl[\prod_{i=1}^l
 \int\mathrm{d}^3s_i\,\sum_{B_i}n_{B_i}(\mathbf{s}_i)\Biggr]
 U_{AB_1\ldots B_l}
 (\mathbf{r}_A,\mathbf{s}_1,\ldots,\mathbf{s}_l),
\end{equation}
where
\begin{multline}
\label{Eq63}
U_{A_1\ldots A_j}(\mathbf{r}_{1},\ldots,\mathbf{r}_{j})
 =\frac{(-1)^{j-1}\hbar\mu_0^j}
 {(1+\delta_{2j})\pi}\int_0^\infty\mathrm{d}u\,u^{2j}
 \alpha_{A_1}(iu)\cdots\alpha_{A_j}(iu)\\
\times\mathcal{S}\mathrm{Tr}\bigl[
 \overline{\mathbf{H}}(\mathbf{r}_1,\mathbf{r}_2,iu)\cdots
 \overline{\mathbf{H}}(\mathbf{r}_j,\mathbf{r}_1,iu)\bigr]
\end{multline}
is nothing but the $j$-atom vdW potential on an arbitrary
magnetodielectric background
$\overline{\varepsilon}(\mathrm{r},\omega)$,
$\mu(\mathbf{r},\omega)$.

We have hence proved that the CP interaction of an atom
with a macroscopic dielectric body which is described within the
framework of macroscopic QED is the result of all possible
microscopic many-atom vdW interactions between the atom under
consideration and the atoms forming the body, provided that the
susceptibility is of Clausius-Mosotti type (\ref{Eq48}) and the
convergence condition (\ref{Eq58}) holds -- generalising the
result in Ref.~\cite{Milonni92}. Conversely, our proof shows that when
Eqs.~(\ref{Eq62}) and (\ref{Eq63}) hold and the convergence
condition (\ref{Eq58}) is satisfied, then the electric susceptibility
must have the Clausius-Mosotti form. In addition, the proof has
delivered the general many-atom vdW potentials (\ref{Eq63}) on an
arbitrary magnetodielectric background. For $j$ $\!=$ $\!2$,
Eq.~(\ref{Eq63}) agrees with the two-atom potential (\ref{Eq37})
derived in Sec.~\ref{Sec3.2}, while for higher $j$, it presents a
generalisation of the free-space vdW potentials derived in
Ref.~\cite{Power85}.

The applicability of the microscopic expansion (\ref{Eq62}) depends
crucially on the validity of the convergence condition (\ref{Eq58}).
Recalling Eq.~(\ref{Eq20}) and estimating
\begin{equation}
\label{Eq64}
\sum_Bn_B(\mathbf{r})\,\frac{\alpha_B(0)}{\varepsilon_0}
=\sum_Bn_B(\mathbf{r})\,\frac{2}{3\hbar\varepsilon_0}\sum_l
 \frac{|\mathbf{d}^{0l}_B|^2}{\omega_B^l}
\approx\frac{1}{V_\mathrm{s}}\,
 \frac{2q_\mathrm{e}^2a_\mathrm{B}^2}{3\varepsilon_0E_\mathrm{R}}\,f
=\frac{4fV_\mathrm{A}}{V_\mathrm{s}}
\end{equation}
[$q_\mathrm{e}$, electron charge; $m_\mathrm{e}$, electron mass;
$a_\mathrm{B}$ $\!=$ $\!\hbar/(\alpha_0m_\mathrm{e}c)$;
$\alpha_0$ $\!=$ $\!q_\mathrm{e}^2/(4\pi\varepsilon_0\hbar
c)$; $E_\mathrm{R}$ $\!=$ $\!\hbar^2/(2m_\mathrm{e}a_\mathrm{B}^2)$;
$V_\mathrm{s}$, volume accessible per atom within the body;
$V_\mathrm{A}$ $\!=$ $\!4\pi a_\mathrm{B}^3/3$, \mbox{$f$ $\!>$ $\!1$},
species-dependent factor], Eq.~(\ref{Eq58}) can be reformulated as
$V_\mathrm{s}$ $\!\gtrsim$ $\!8fV_\mathrm{A}/3$, stating simply that
the atoms must be well-separated within the body. If this is not the
case, the microscopic expansion (\ref{Eq62}) does not converge, while
the more general macroscopic expression (\ref{Eq19}) for the CP
potential remains valid.


\section{Summary}
\label{Sec5}

We have demonstrated that on the basis of macroscopic QED in linear,
causal media, leading-order perturbation theory can be employed to
derive general expressions for both the single-atom CP potential and
the two-atom vdW potential in the presence of an arbitrary
arrangement of magnetodielectric bodies.

Moreover, starting from this very general, geometry-independent basis,
we have used the Born expansion of the Green tensor together with the
Clausius-Mosotti law to prove that the CP interaction of a single
atom with inhomogeneous, dispersing and absorbing dielectric bodies in
the presence of an arbitrary magnetodielectric background can be
written as a sum of many-atom vdW potentials. The proof demonstrates
the equivalence of the microscopic and macroscopic descriptions
provided that the microscopic picture is applicable, while at the same
time delivering explicit expressions for the general many-atom vdW
potentials in the presence of magnetodielectric media.


\section*{Acknowledgements}
This work was supported by the Deutsche Forschungsgemeinschaft. H.S.
would like to thank the Ministry of Science, Research, and Technology
of Iran. H.T.D. would like to thank the Alexander von Humboldt
Stiftung and the National Program for Basic Research of Vietnam. 


\end{document}